\documentclass[fleqn,usenatbib]{mnras}
\usepackage{newtxtext,newtxmath}
\usepackage[T1]{fontenc}
\DeclareRobustCommand{\VAN}[3]{#2}
\let\VANthebibliography\thebibliography
\def\thebibliography{\DeclareRobustCommand{\VAN}[3]{##3}\VANthebibliography}

\usepackage{booktabs}
\usepackage{ae,aecompl}
\usepackage{xcolor}
\usepackage{float}
\usepackage{graphicx}
\usepackage{subfigure}
\usepackage{amsmath}	
\usepackage{tabularx}

\title[]{On the Precision of Full-spectrum Fitting of Stellar Populations. III. Identifying Age Spreads.}

\author[Asa'd et al.]{Randa Asa'd,$^{1,2}$\thanks{E-mail:raasad@aus.edu}
Paul Goudfrooij,$^{2}$
A. M. As'ad,$^{3}$
H. G. El-Mir,$^{4}$
L. Begum,$^{4}$
\newauthor
A. Aljasmi,$^{4}$
O. Almatroushi$^{4}$\smallskip \\ 
$^{1}$American University of Sharjah, Physics Department, P.O.Box 26666, Sharjah, UAE\\
$^2$Space Telescope Science Institute, 3700 San Martin Drive, Baltimore, MD 21218, USA\\
$^3$University of Jordan, Amman, Jordan\\
$^{4}$American University of Sharjah, P.O.Box 26666, Sharjah, UAE
}

\begin{document}

\pagerange{\pageref{firstpage}--\pageref{lastpage}} \pubyear{2020} 

\maketitle

\label{firstpage}

\begin{abstract}

In this third paper of a series on the precision of obtaining ages of stellar populations using the full spectrum fitting technique, we examine the precision of this technique in deriving possible age spreads within a star cluster. We test how well an internal age spread can be resolved as a function of cluster age, population mass fraction, and signal-to-noise (S/N) ratio. For this test, the two ages (Age\,(SSP$_1$) and Age\,(SSP$_2$)) are free parameters along with the mass fraction of SSP$_1$. 
We perform the analysis on 118,800 mock star clusters covering all ages in the range 6.8 $<$ log\,(age/yr) $<$ 10.2, with mass fractions from 10\% to 90\% for two age gaps (0.2 dex and 0.5 dex). Random noise is added to the model spectra to achieve S/N ratios between 50 to 100 per wavelength pixel. 
We find that the mean of the derived Age\,(SSP$_1$) generally matches the real Age\,(SSP$_1$) to within 0.1 dex up to ages around log (age/yr) = 9.5. The precision decreases for log (age/yr) $>$ 9.6 for any mass fraction or S/N, due to the similarity of SED shapes for those ages. In terms of the recovery of age spreads, we find that the derived age spreads are often larger than the real ones, especially for log(age/yr) $\la$ 8.0 and high mass fractions of SSP$_1$.  Increasing the age gap in the mock clusters improves the derived parameters, but Age\,(SSP$_2$) is still overestimated for the younger ages.

\end{abstract}


\section{Introduction}
\label{Introduction}

In the first paper of this series \citep[][hereafter paper I]{Asad20}, we investigated the precision of the ages and metallicities of 21,000 mock simple stellar populations (SSPs) determined through full-spectrum fitting in the optical range. We found that for S/N $\geq$ 50, this technique can yield ages of SSPs to an overall precision of $\Delta\,\mbox{log(age/yr)} \sim 0.1$ for ages in the ranges 7.0 $\leq$ log\,(age/yr) $\leq$ 8.3 and 8.9 $\leq$ log\,(age/yr) $\leq$ 9.4. For the age ranges of 8.3 $\leq$ log\,(age/yr) $\leq$ 8.9 and log\,(age/yr) $\geq$ 9.5,  the age uncertainty rises to about $\pm 0.3$ dex. In the second paper of the series (\citealt{Goudfrooij21}, hereafter paper II) we studied the influence of star cluster mass through stochastic fluctuations of the number of stars near the top of the stellar mass function, which dominate the flux in certain wavelength regimes depending on the age. Studying several wavelength intervals in the 0.35\,--\,5.0 $\mu$m range, we found that in general, the spectral shape in the blue optical regime (0.35\,--\,0.7\,$\mu$m) is by far the least impacted by stochastic fluctuations among the wavelength intervals studied.
In both previous studies we followed the assumption that star clusters as SSPs, considering all constituent stars to have the same age and metallicity. However, most old globular clusters are now known to host abundance variations of several light elements \citep[mainly He, C, N, O, Na, and Al; see][for details]{BastianLardo18}. This phenomenon is often referred to as Multiple Stellar Populations (MSPs).   
Interestingly, MSPs have not yet been identified in clusters younger than $\sim$\,2 Gyr \citep[see][]{Martocchia18, Asad_ESO2020, Cabrera-Ziri20, Li20}. However, if young massive clusters (YMCs) are the young analogs of Globular Clusters (GCs), one might expect to find MSPs in YMCs. The role that age plays in the MSP phenomenon has been difficult to constrain \citep[e.g.,][]{Cabrera-Ziri20}. 
Additionally, several studies, based on accurate Hubble Space Telescope (HST) photometry and Gaia have revealed that the color-magnitude diagram (CMD) of stellar clusters younger than $\sim$\,2 Gyr exhibit an extended main sequence turn off (eMSTO) which is not due to photometric errors, field-star contamination, differential reddening, or non-interacting binaries \citep[e.g.][and references therein]{Bertelli03, Mackey07, Milone09,Goudfrooij09, Cordoni18}. These findings have challenged the traditional picture that the CMDs of star clusters are well described by SSPs, and have triggered huge efforts to understand eMSTOs.
The eMSTO phenomenon was initially proposed to be due to multiple stellar formation events, challenging the conventional belief that star clusters are groups of coeval stars. However,  alternative scenarios to explain this phenomenon were soon proposed as well. The effects of stellar rotation, first proposed by \citet{Bastian09}, is currently one of the most widely accepted scenarios \citep[see, e.g.,][]{dupree2017a, marino2018b, Bastian18, Sun19}.
However, it is still argued that age spreads may be present in clusters \citep[see][]{Goudfrooij17,Goudfrooij18b,Gossage19,Costa19}. 

Integrated-light spectra of star clusters have proven to be accurate tools for studying extragalactic clusters for which spatially resolved data is not available \citep{AA02, Puzia05,  Puzia06, Santos06, Palma08, Talavera10, Fernandes10, Asad13, Asad14,  Chilingarian18}. In this work we aim to address the question: To what extent can integrated-light spectra be used to identify age spreads at different ages?

\section{Mock Star Clusters}

We create a sample of mock star cluster spectra following the results of Paper I: A) We used the wavelength range $3700-5000$\,\AA\ which we showed to be optimal for deriving accurate ages from integrated-light spectra using the full-spectrum fitting method.   B) our sample has S/N $>$ 50; C) the smallest age difference (gap) used to create the age combinations is 0.2 dex.  
We use the flexible stellar population synthesis (FSPS) code \citep{Conroy09, Conroy10} operated through the Python package python-FSPS \citep{python-fsps} to create integrated-light spectra of SSPs for the wavelength range $3700-5000$\,\AA\ with \citet{Kroupa01} IMF and fixed metallicity [Z$/$H] = $-$0.4 for the age range 7.0 $<$ log age $<$ 10.2 in steps of 0.1 dex) using MIST isochrones \citep{Choi17} and the MILES spectral library \citep{MILESI, Vazdekis10, Vazdekis16}, which has a spectral resolution of 2.5 \AA   \citep{Falcon11}.  We use the default settings of python-FSPS (similar to Paper I and II), including the absence of nebular emission for the youngest ages, and a zero fraction of blue horizontal branch stars at old ages.
We then combine two SSPs, in order to explore if our method can derive age spreads where two SSPs are present, according to the equation:
\begin{equation} 
f\,SSP_{1}+(1-f)\,SSP_{2}
\label{Eq_2SSPs}
\end{equation}  
\noindent where $f$ is the fractional contribution of $SSP_{1}$ by mass, running from 0 to 1 in steps of 0.1, $SSP_{1}$ is the SSP of an arbitrary age, and $SSP_{2}$ is the additional SSP. 
We start with a sample that has an age gap of 0.2 dex and can be described by the following equation:
\begin{equation} 
\log \mbox{(age(SSP$_{2}$))} = \log \mbox{(age(SSP$_{1}$))} + 0.2. 
\end{equation} 
Finally, we add 30 different realizations of random noise, corresponding to S/N values between 50 to 100 (in steps of 10) for each combination, thus producing a grand total of 59,400 mock clusters. 

\section{Fitting Single Age}
\subsection{Method}

We first assume that the sample of mock clusters consist of only one SSP, and use the full-spectrum fitting program {\sc ASAD}$_{\rm 2}$
 \citep[for a full description, see][]{Asad13,Asad14,Asad16} to obtain the best-fitting age using the following equation:  
\begin{equation} 
\sum_{\lambda=\lambda_{\rm initial }}^{\lambda_{\rm final}}
\frac{[(OF)_{\lambda} - (MF)_{\lambda}]^{2}}{(OF)_{\lambda_{\rm norm}}}.
\label{Eq_1SSP}
\end{equation} 
where OF is the mock cluster flux, MF is the SSP model flux, and
$\lambda_{\rm norm}$ is the wavelength at which the model and the mock cluster spectra are normalized.

\subsection{Results}

When $f$ = 0 in equation \ref{Eq_2SSPs}, the mass fraction of SSP$_1$ = 0, the spectrum is a single SSP (100\% contribution from SSP$_2$). In this case the correct age is always correctly derived for all S/N.
\begin{figure*}
\resizebox{185mm}{!}{\includegraphics[angle=270]{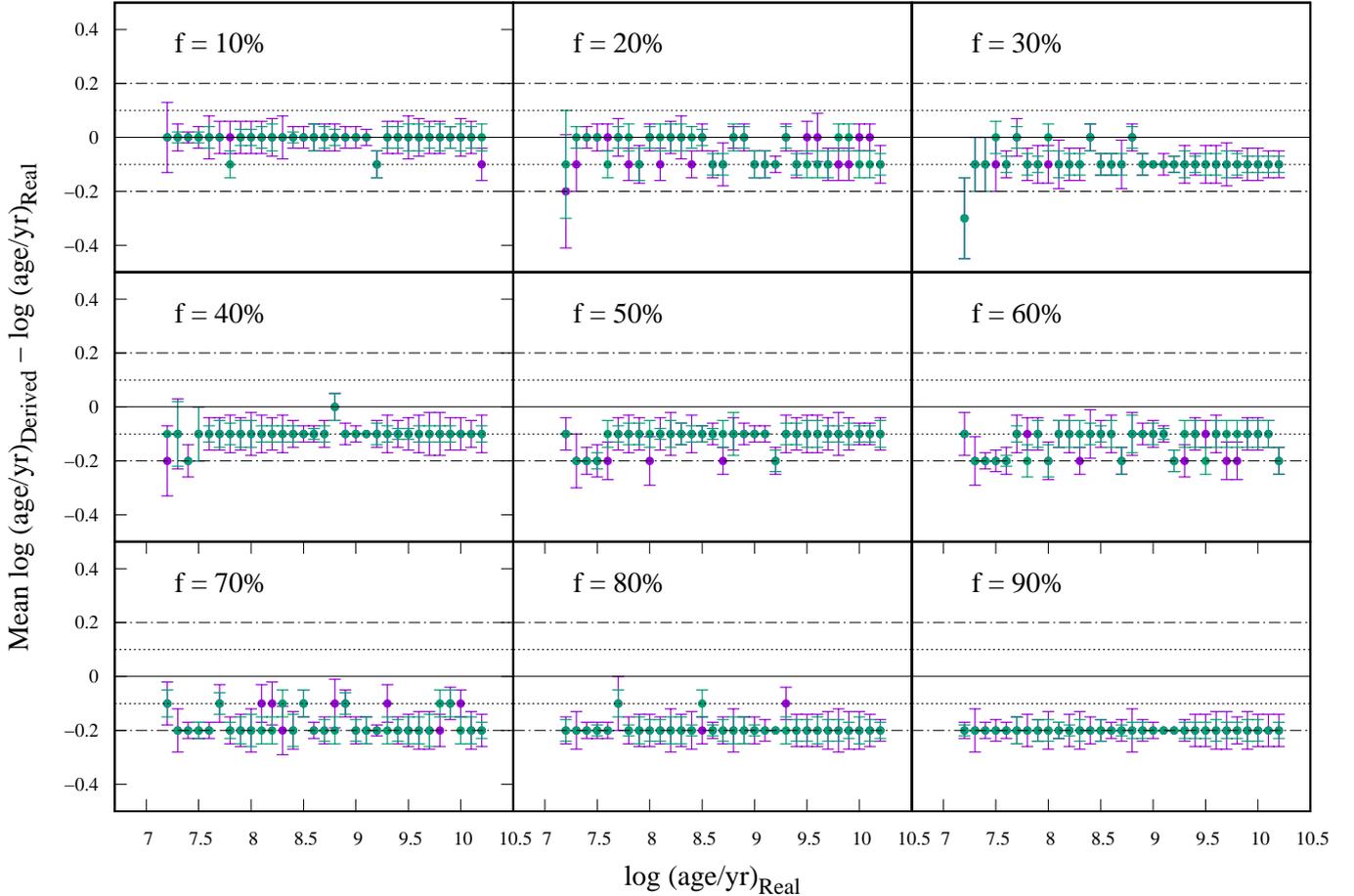}}
\caption{The mean derived single age and the uncertainty (calculated as the standard deviation) for the 30 realization of each set versus real age. The green dots represent results for S/N = 100 and the purple dots represent the results for S/N = 50. The continuous middle line is the y = 0 line, the dotted lines are y = $\pm$ 0.1 and the dashed lines are y = $\pm$ 0.2}
\label{SingleAge}
\end{figure*}
The results obtained for $0 < f < 1$ are shown in Figure \ref{SingleAge}, where we plot the mean of the derived single age with the uncertainty (calculated as the standard deviation) for the 30 realizations of each S/N and mass fraction $f$ versus the input age of SSP$_2$ in equation 1. \\
The figure shows that as the mass fraction of SSP$_1$ increases, the full-spectrum fitting technique increasingly favours the average age values [i.e., ((SSP$_1$)+(SSP$_2$))/2]. Once $f > 60$\%, where SSP$_1$ dominates, the full-spectrum fitting technique increasingly derives the age of SSP$_1$ correctly. For all ages and mass fraction values, there is no significant dependence on the S/N value in the range tested.

\section{Fitting Two Ages}
\subsection{Method}

We create the two-age model grid, using equation 1 for all possible combinations of SSP$_1$ and SSP$_2$ with an age gap of 0.2 dex between the two SSPs, and attempt to recover Age(SSP$_1$), Age(SSP$_2$) and $f$ using the full-spectrum fitting technique.  The results obtained are arranged so that Age(SSP$_1$) is always less than Age(SSP$_2$). We tested the technique first on the trivial case where no noise was added to the combined spectra and the input parameters were perfectly recovered.

\subsection{Results}

\subsubsection {Derived Age(SSP$_1$) and Age(SSP$_2$)}
\begin{figure*}
\resizebox{165mm}{!}{\includegraphics[angle=270]{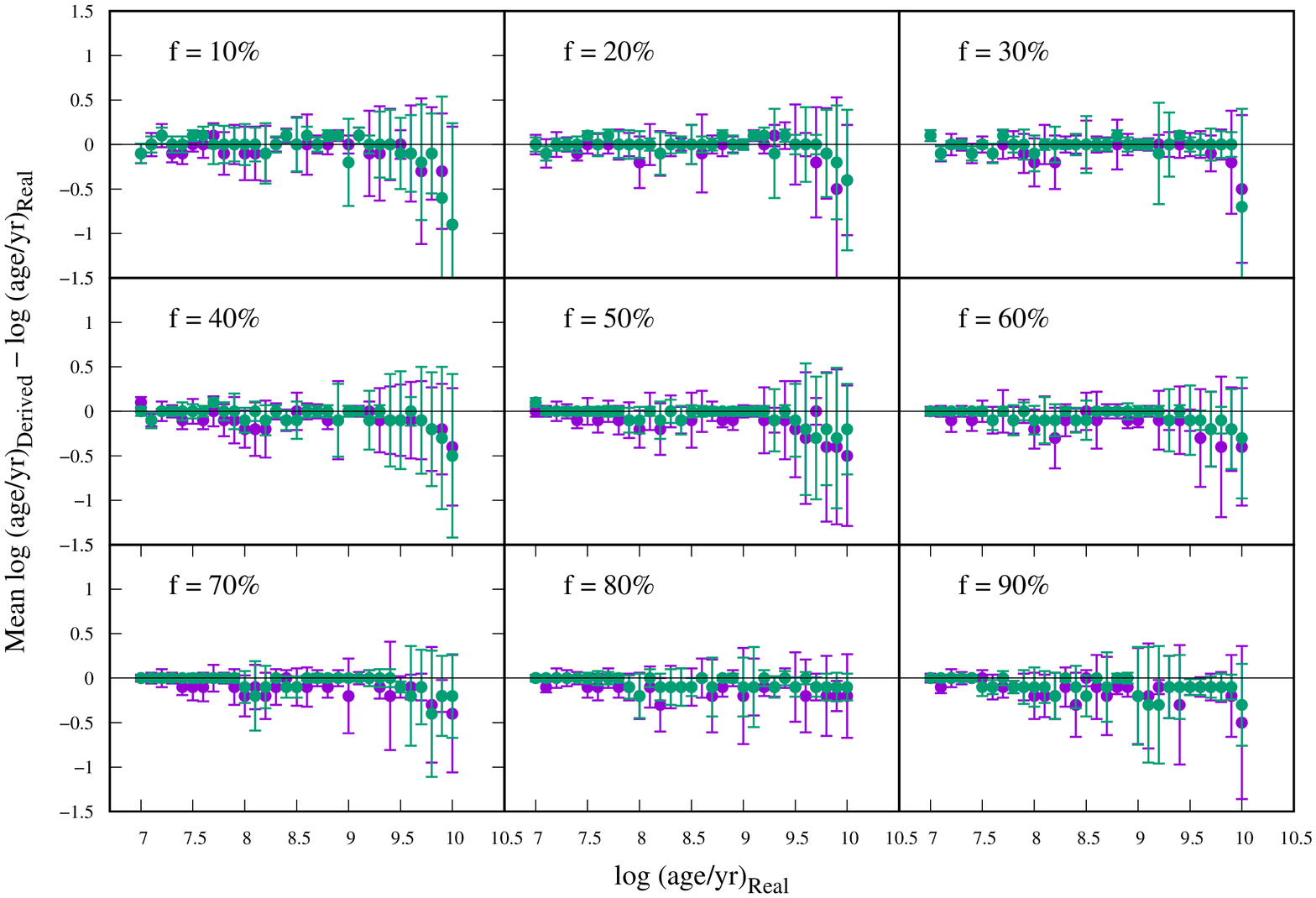}}
\resizebox{165mm}{!}{\includegraphics[angle=270]{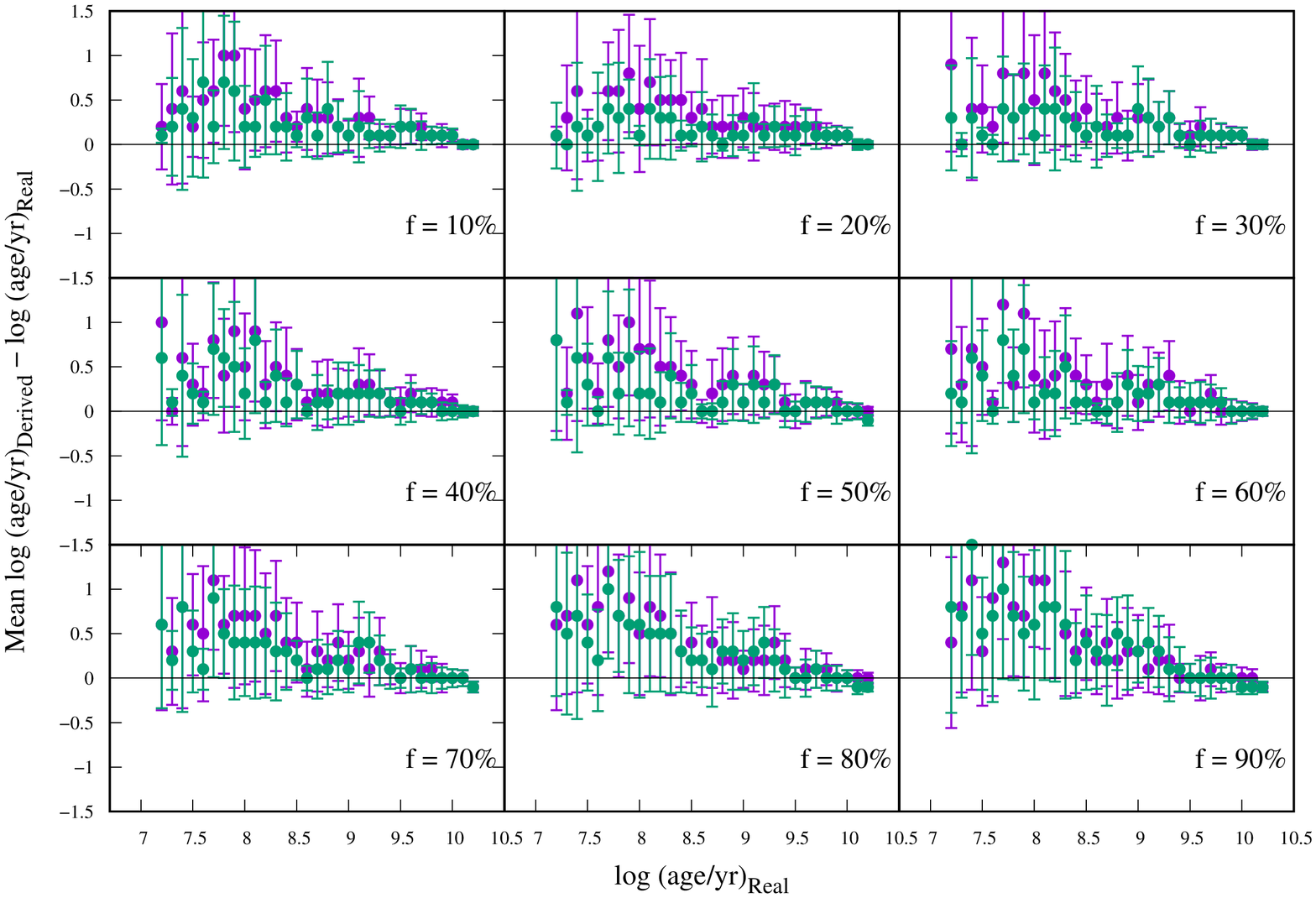}}
\caption{The mean of the derived age versus the input age. The top panel shows the results of Age(SSP$_1$) and the bottom panel shows the results of Age(SSP$_2$).The green dots represent results for S/N = 100 and the purple dots represent the results for S/N = 50. }
\label{2SSPs_ss2}
\end{figure*}

In order to assess to what extent integrated-light spectra can be used to identify age spreads at different ages, we first examine the mean derived age for each input SSP age in the combination.
Figure \ref{2SSPs_ss2} shows the mean of the derived ages with the uncertainty (calculated as the standard deviation) for the 30 realizations of each S/N and mass fraction $f$ versus the input age. 
The top set of panels shows the results of Age(SSP$_1$) and the bottom set of panels shows the results of Age(SSP$_2$). 
Overall, the mean of the derived Age(SSP$_1$) matches the real Age(SSP$_1$) up to ages around log (age/yr) = 9.5. The precision decreases for log (age/yr) $>$ 9.6 for any mass fraction or S/N. 
The shape of the
SED does not change much beyond log\,(age/yr) = 9.6, causing age estimates to be relatively imprecise for those ages (see Paper I and Paper II for more discussion on this).  

The uncertainty associated with the mean of the derived Age(SSP$_2$) (see lower set of panels of Figure \ref{2SSPs_ss2}) is much higher than that of Age(SSP$_1$), especially for young ages. To shed more light on the cause of this result, we list the results of the 30 realizations with S/N = 100 obtained for the input combination of 80\% of log(Age(SSP$_1$)) = 7.0 plus 20\% of log(Age(SSP$_2$)) = 7.2 in Table \ref{T1}.
The list shows that the full-spectrum fitting technique tends to derive combinations of a young age that is close to the real age along with an older age, with the latter age encompassing several orders of magnitude from one mock cluster to another. This is found to be generally the case for young ages (log (age/yr) $<$ 8.0) and is explained by the fact that SEDs for log (age/yr) in the range 7.0\,--\,7.5 are very similar to one another, and they produce significantly more flux than older populations (especially in the blue; see, e.g., Papers I and II). Hence, the derived two-SSP combination includes one young SSP that is close to the real age, with an older SSP with almost \textit{any} other age, since the latter doesn't produce a significant percentage of the flux.

\begin{table}

  \caption{Derived results for the input combination of 80\% Age(SSP$_1$) = 7.0 with 20\% Age(SSP$_2$) = 7.2}

 \begin{tabular}{ |lll|lll|lll|}
\toprule
\multicolumn{1}{|l}{Age$_{1}$} &
\multicolumn{1}{l}{$f$} & \multicolumn{1}{l}{Age$_{2}$} & \multicolumn{1}{|l}{Age$_{1}$} & \multicolumn{1}{l}{$f$} &
\multicolumn{1}{l}{Age$_{2}$}  & \multicolumn{1}{|l}{Age$_{1}$} & \multicolumn{1}{l}{$f$} &
\multicolumn{1}{l|}{Age$_{2}$}\\
\cmidrule(lr){1-1}\cmidrule(lr){2-2}\cmidrule(lr){3-3}\cmidrule(lr){4-4}\cmidrule(lr){5-5}\cmidrule(lr){6-6}\cmidrule(lr){7-7}\cmidrule(lr){8-8}\cmidrule(lr){9-9}

6.8 & 10\% & 7.0 & 7.0 & 80\% & 7.3 & 7.0 & 70\% & 7.8 \\
7.0 & 40\% & 7.1 & 7.0 & 80\% & 7.3 & 7.0 & 40\% & 8.9 \\
7.0 & 40\% & 7.1 & 7.0 & 70\% & 7.3 & 7.0 & 40\% & 9.0 \\
7.0 & 30\% & 7.1 & 7.0 & 60\% & 7.3 & 7.0 & 30\% & 9.1 \\
7.0 & 80\% & 7.2 & 7.0 & 70\% & 7.3 & 7.0 & 20\% & 9.4 \\
7.0 & 80\% & 7.2 & 7.0 & 70\% & 7.3 & 7.0 & 20\% & 9.4 \\
7.0 & 80\% & 7.2 & 7.0 & 70\% & 7.6 & 7.0 & 20\% & 9.4 \\
7.0 & 80\% & 7.2 & 7.0 & 70\% & 7.7 & 7.0 & 10\% & 9.8 \\
7.0 & 80\% & 7.2 & 7.0 & 70\% & 7.7 & 7.0 & 10\% & 9.8 \\
7.0 & 80\% & 7.3 & 7.0 & 70\% & 7.7 & 7.0 & 10\% & 9.8 \\
\toprule
\label{T1}
 \end{tabular}
\end{table}

\begin{figure*}
\resizebox{120mm}{!}{\includegraphics[angle=0]{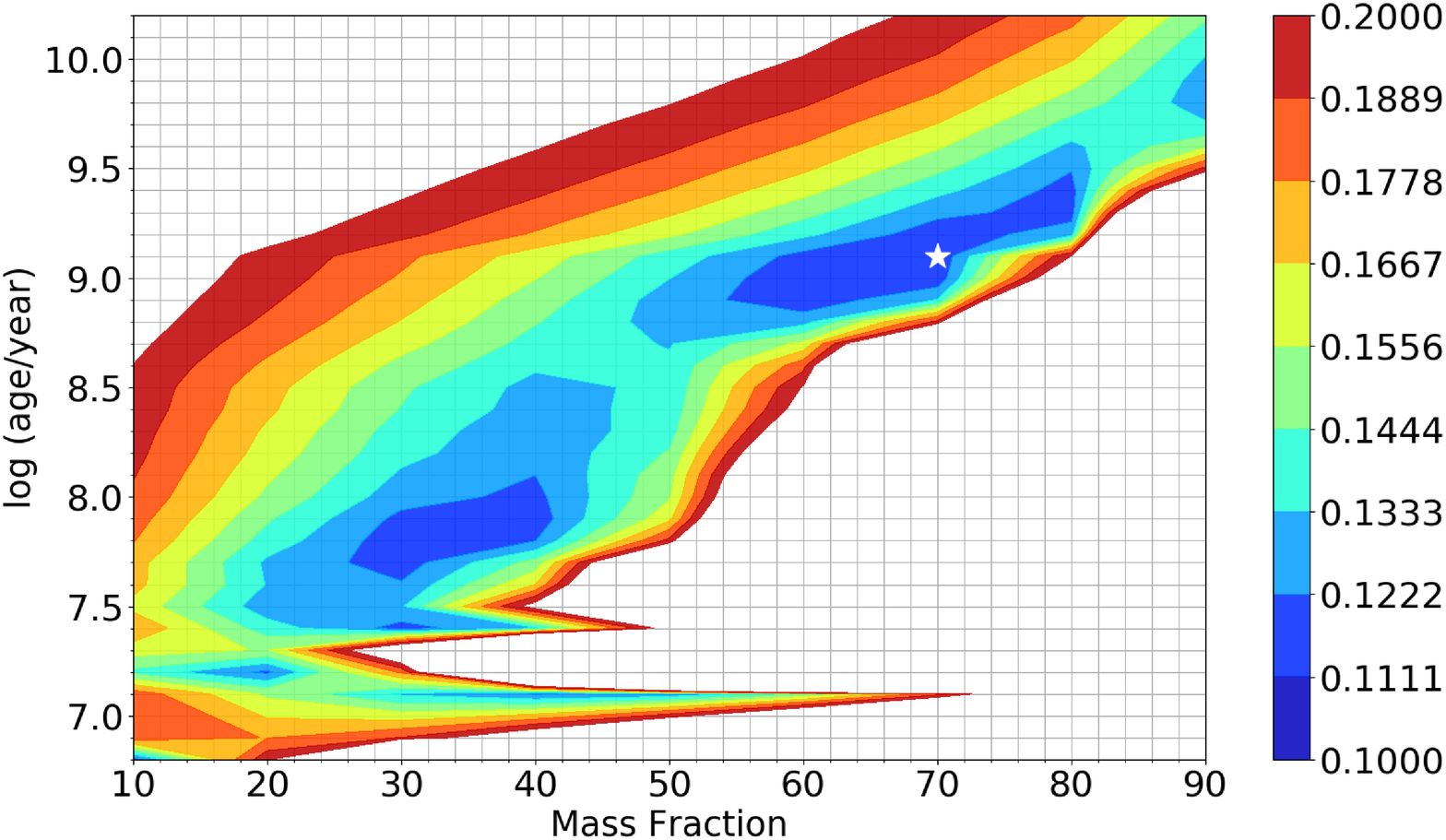}}
\caption{The heat map for Age(SSP$_2$) of the mock cluster where the derived parameters are Age(SSP$_1$) = 7.0 with 70\% Age(SSP$_2$) = 9.1 in table \ref{T1}. The white star shows the coordinates of the combination with the minimum $\chi^2$ value.}
\label{contours}
\end{figure*}

To illustrate this further, a closer look at the mock cluster for which the derived parameters are 30\% of Age(SSP$_1$) = 7.0 with 70\% Age(SSP$_2$) = 9.1 reveals that the $\chi^2$ value of this combination is 0.11212. Figure~\ref{contours} shows a heat map with all the combinations of Age(SSP$_2$) and mass fraction that have $\chi^2$ values between 0.1 and 0.2. The white star shows the coordinates of the combination with the minimum $\chi^2$ value. The dark blue regions represent the $\chi^2$ values between 0.100 and 0.111. This shows that combinations of Age(SSP$_1$) = 7.0 with Age(SSP$_2$) = 7.7 to 7.9 and 8.9 to 9.4 are very close to the correct combination (80\% of Age(SSP$_1$) = 7.0 with 20\% Age(SSP$_2$) = 7.2). 
This explains the results obtained in the lower panel of Figure \ref{2SSPs_ss2} where the mean of the derived Age(SSP$_2$) is overestimated for young ages.

\begin{figure*}
\resizebox{165mm}{!}{\includegraphics[angle=270]{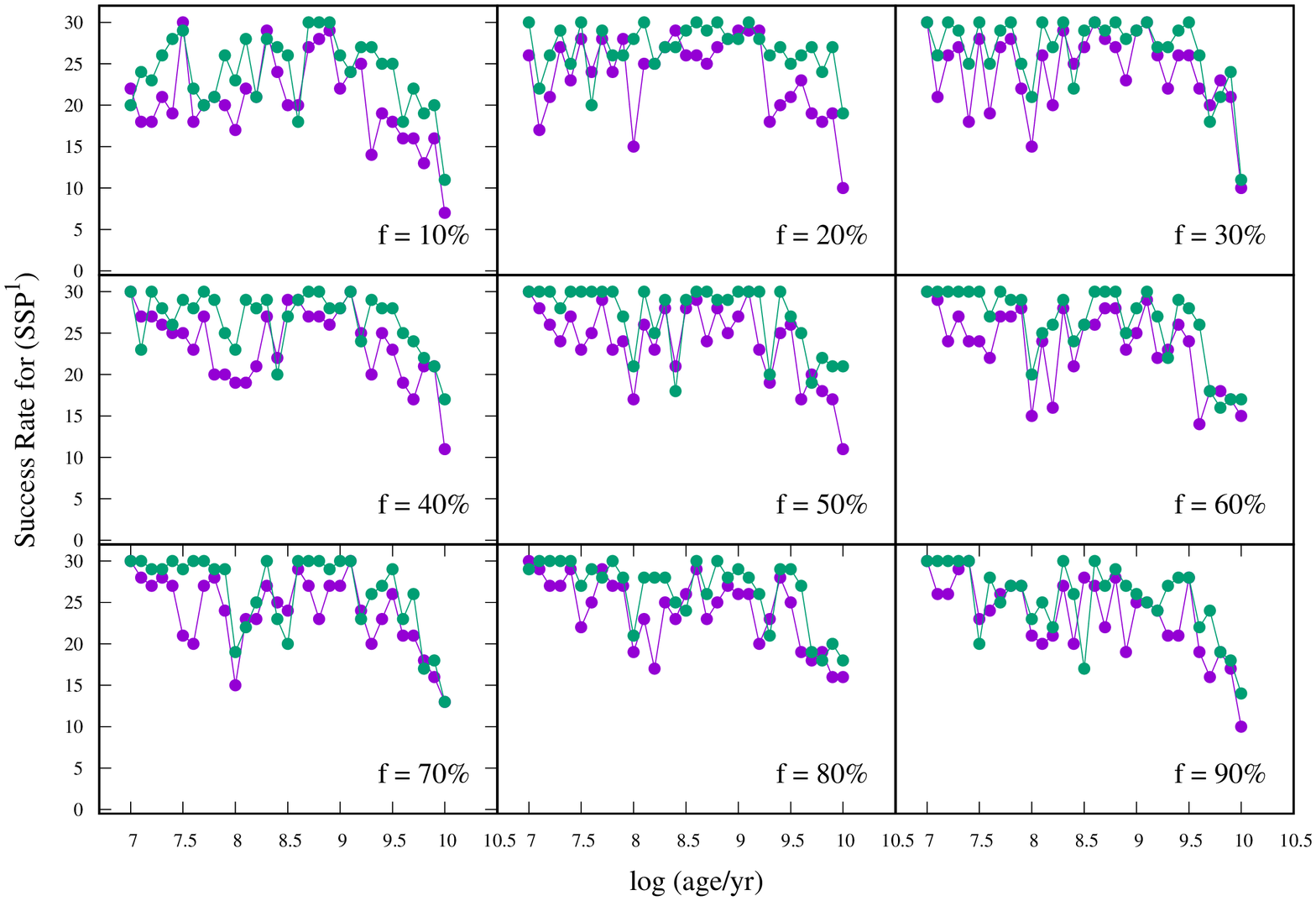}}
\resizebox{165mm}{!}{\includegraphics[angle=270]{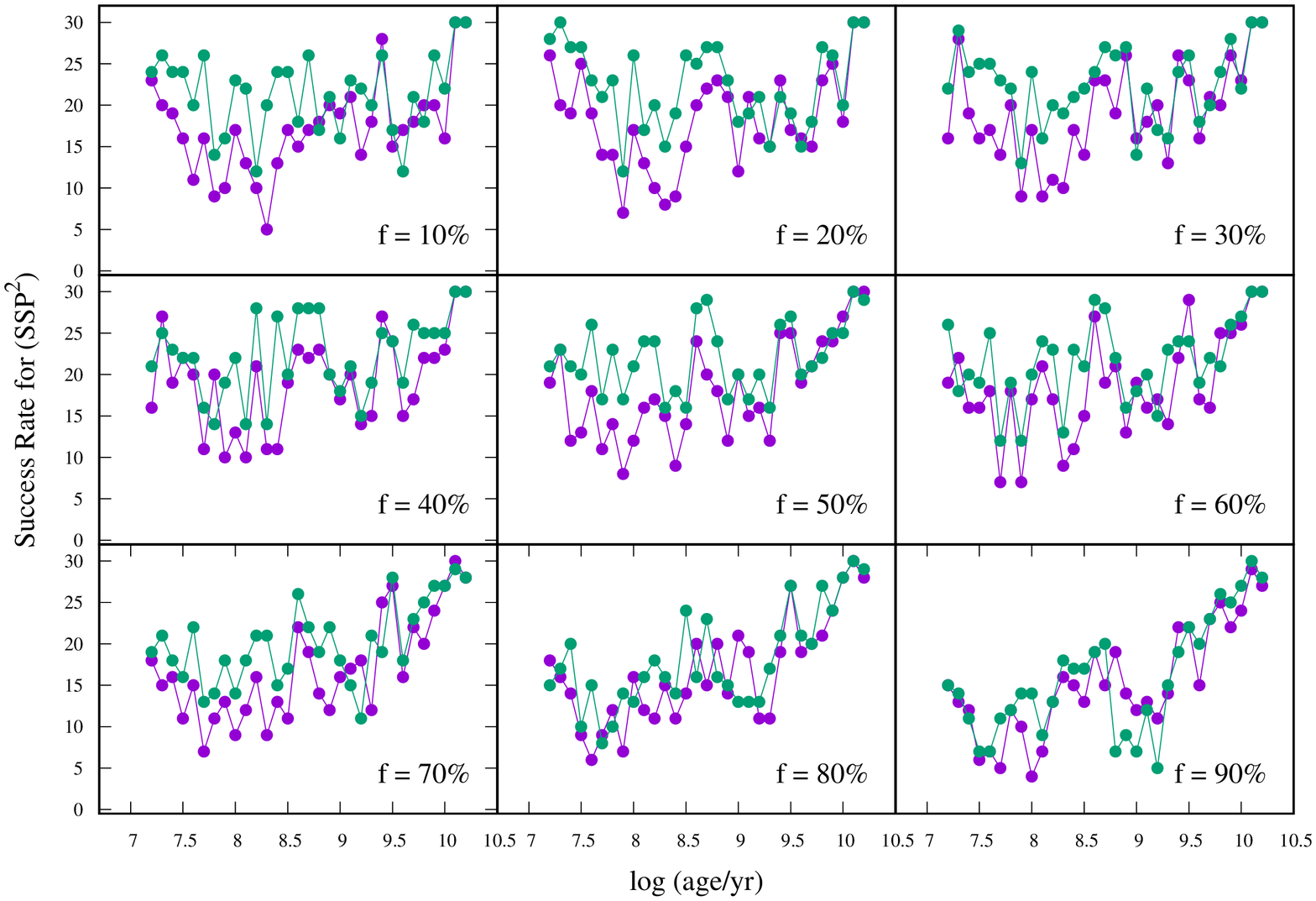}}
\caption{the number of successfully recovered ages (i.e., the number of times the derived age matches the real age within $\pm$ 0.1 dex) as a function of real age. The top panel shows the results of Age(SSP$_1$) and the bottom panel shows the results of Age(SSP$_2$).}
\label{Recovery}
\end{figure*}

Figure \ref{Recovery} shows the ''success rate'' -- defined as the number of times the derived parameter matches the input parameter within $\pm$ 0.1 dex -- as a function of input age.
The success rate of SSP$_{1}$ increases as the input mass fraction $f$ of this SSP increases. The success rate also depends on the input age. It peaks for log (age/year) $\la$ 7.5 and around 9.0 (the latter corresponding to RGB phase transition) and it drops around 8.6 and for $>$ 9.2 for almost all mass fractions. In Paper I we showed that the number of correctly derived ages drop significantly in these age ranges that correspond to the asymptotic giant branch (AGB) phase transition in stellar evolution, and the red giant branch (RGB) phase transition. The age range $>$ 9.2 is the worst in terms of success rate, because SED's in this age range are very similar to one another. Identifying small age spreads of two populations that do not fall around a fast evolutionary transition is not trivial. 
The success rate of SSP$_{2}$ decreases as the input mass fraction $f$ of SSP$_{1}$ increases, as expected. The success rate peaks at ages very slightly younger than the ones at which it peaks for SSP$_{1}$ in addition to log (age/year) = 10 for $f$ from 10\% to 40\%. It is worth noting that this rise toward the oldest ages is meaningless, since we used $\Delta$ SSP = 0.2, so that all the simulations with log(age(SSP1)) > 9.7 or so can only have log(age(SSP2)) higher than that, thus boosting their success numbers.

\subsubsection{Age Spread}

\begin{figure*}
\resizebox{165mm}{!}{\includegraphics[angle=270]{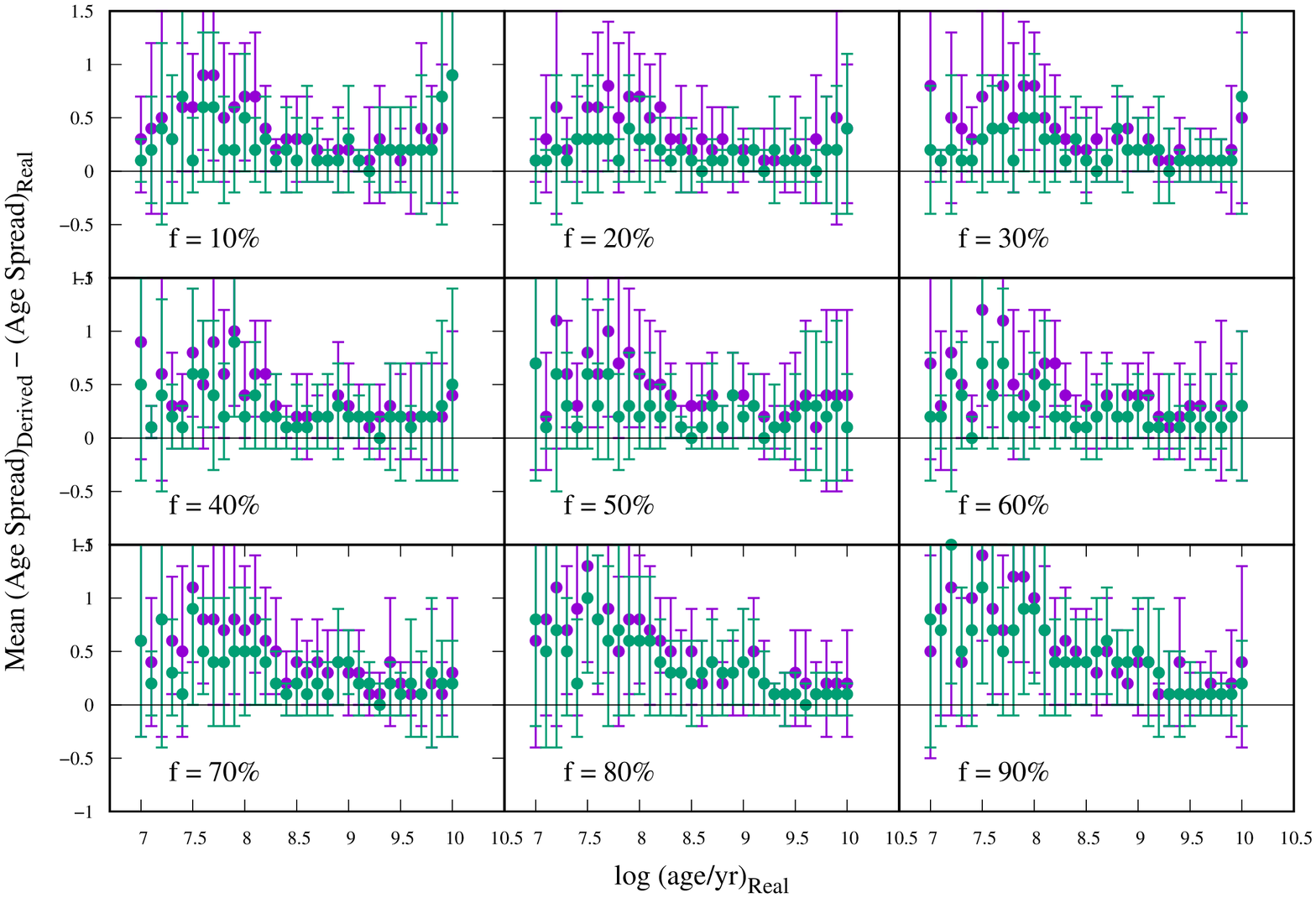}}
\resizebox{165mm}{!}{\includegraphics[angle=270]{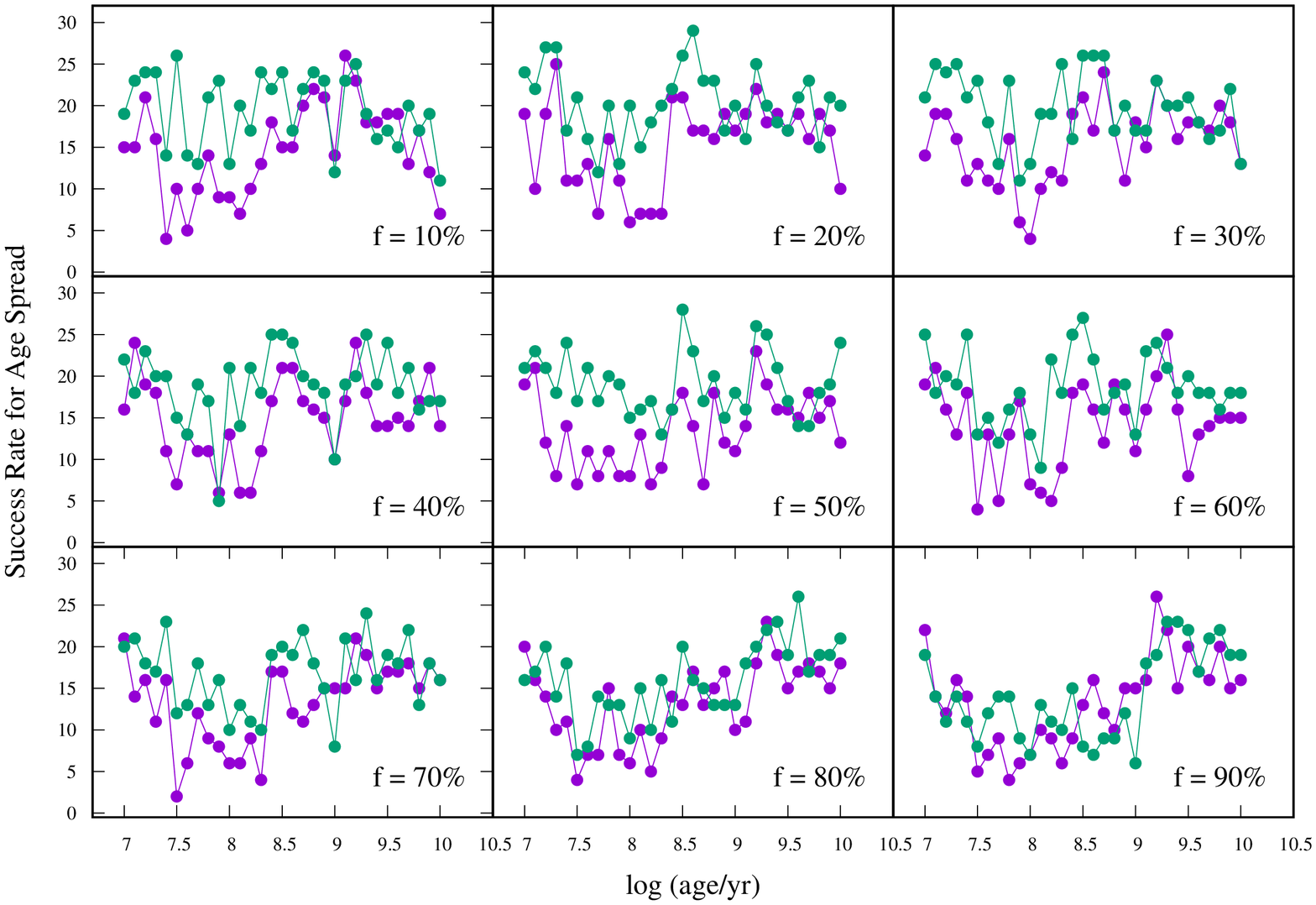}}
\caption{The results for the derived age spread, defined as log\,(Age(SSP$_2$)) $-$ log\,(Age(SSP$_1$)). The top panel shows the mean of the ``derived $-$ input'' age spread as a function of the input age and the lower panel shows the recovery success rate as a function of input age.}
\label{AgeGap_02}
\end{figure*}

Figure \ref{AgeGap_02} shows the results obtained for the age spread, defined as log\,(Age(SSP$_2$)) $-$ log\,(Age(SSP$_1$)). The top panel shows the mean of the ``(derived $-$ input) age spread'' as a function of the input age. The derived average age spreads are larger than the real ones, and the scatter around the mean values is significant, especially for mean log\,(age/yr) $\la$ 8.5. The lower panel shows that the overall success rate for the derived age spread is typically of order 20\%, without a strong dependence on input age or mass fraction.

\section{Increasing the Age Gap}

\begin{figure*}
\resizebox{165mm}{!}{\includegraphics[angle=270]{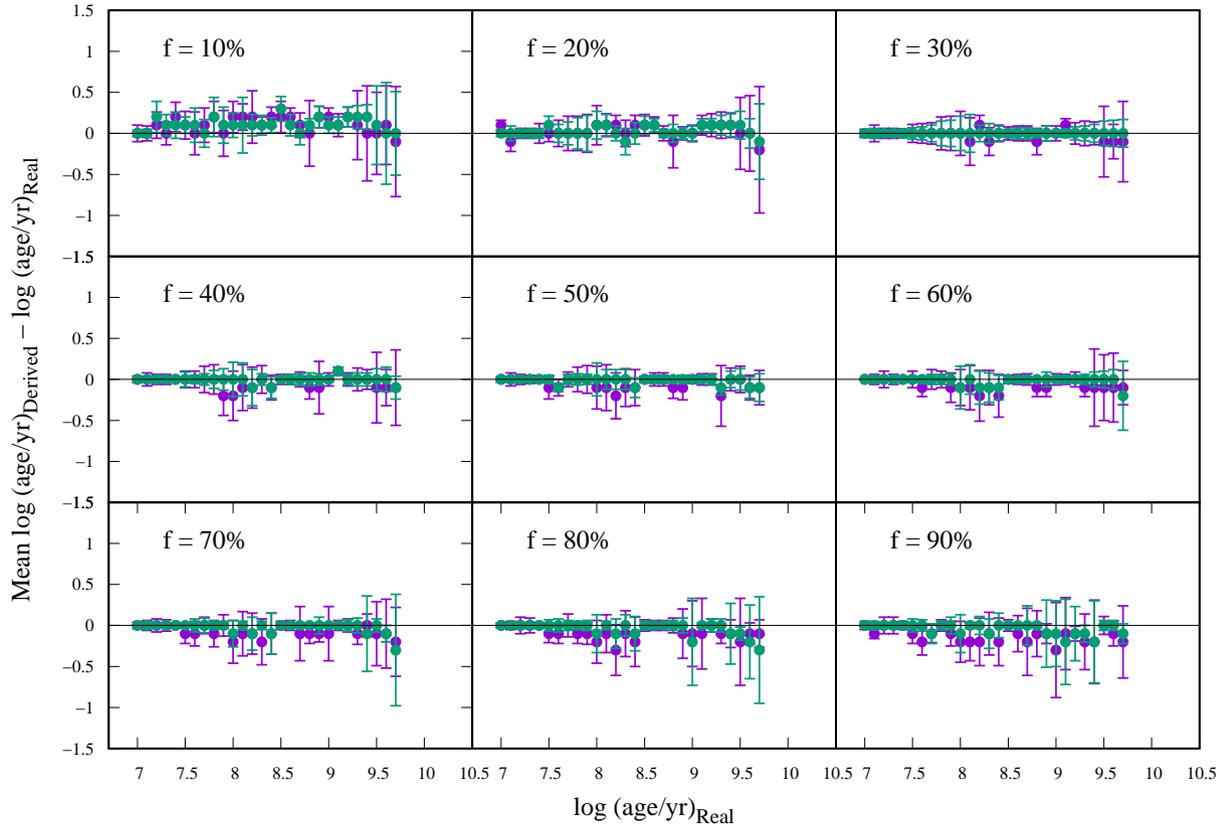}}
\resizebox{165mm}{!}{\includegraphics[angle=270]{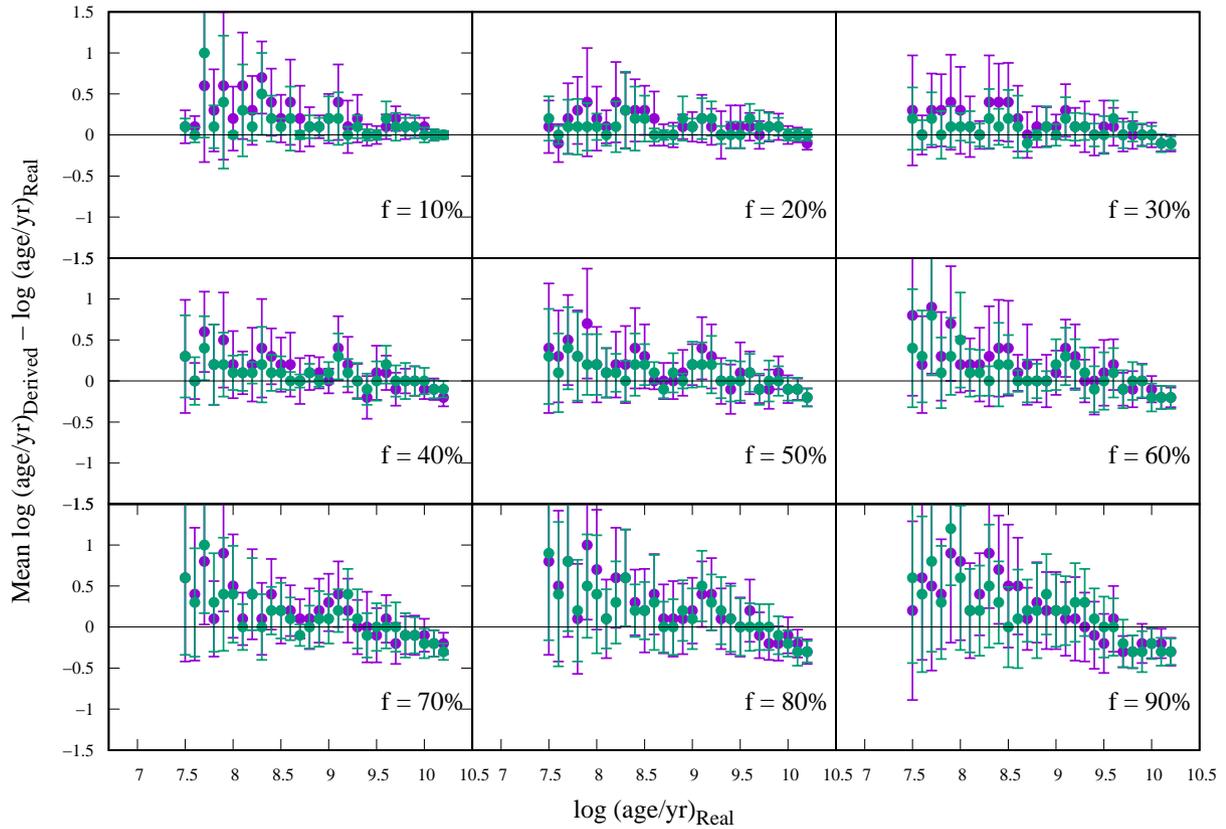}}
\caption{The mean of the derived age versus the input age for mock clusters with Age(SSP$_2$ -  Age(SSP$_1$) = 0.5. The top panel shows the results of Age(SSP$_1$) and the bottom panel shows the results of Age(SSP$_2$). The green dots represent results for S/N = 100 and the purple dots represent the results for S/N = 50. }
\label{Step5_Mean}
\end{figure*}

\begin{figure*}
\resizebox{165mm}{!}{\includegraphics[angle=270]{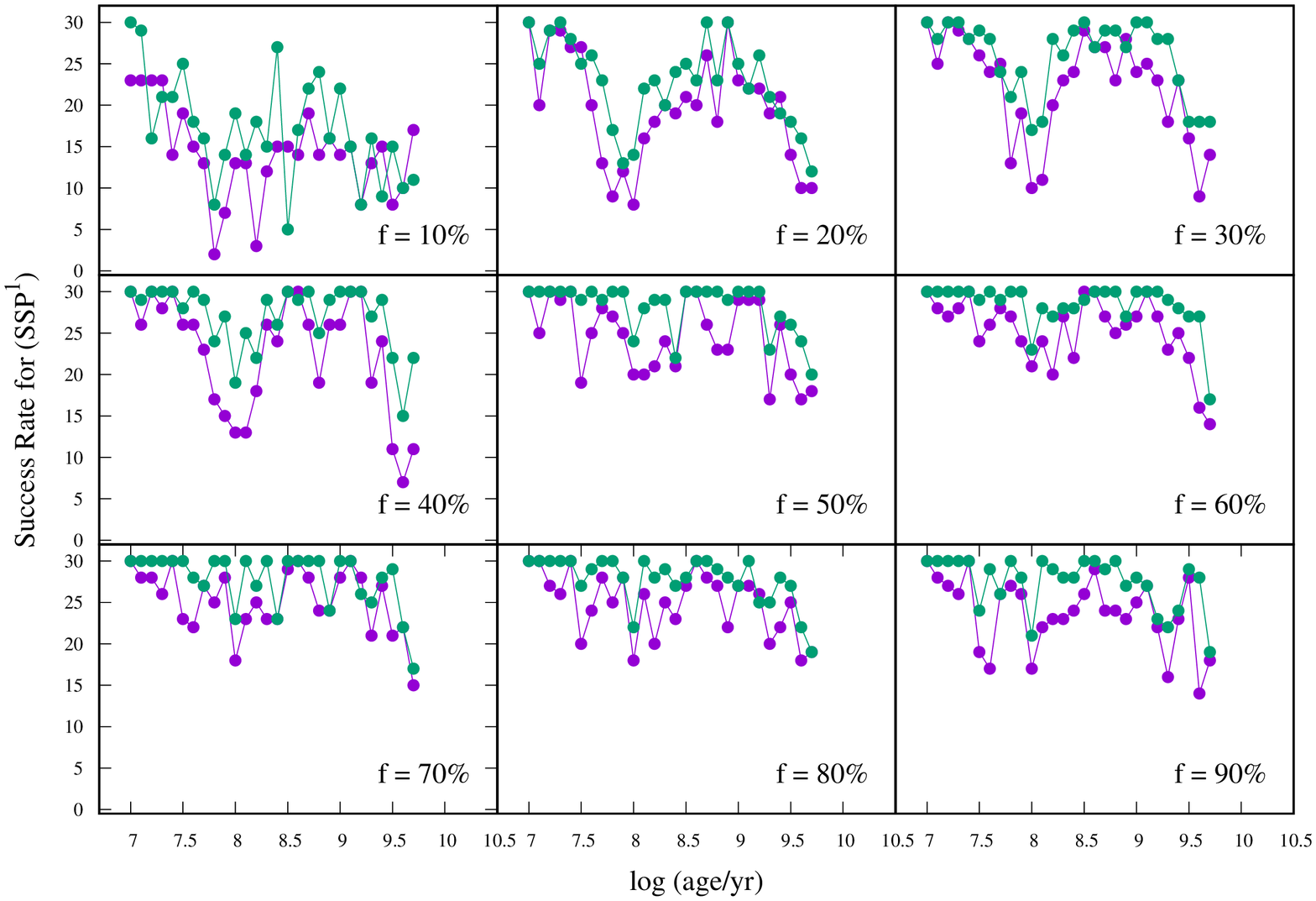}}
\resizebox{165mm}{!}{\includegraphics[angle=270]{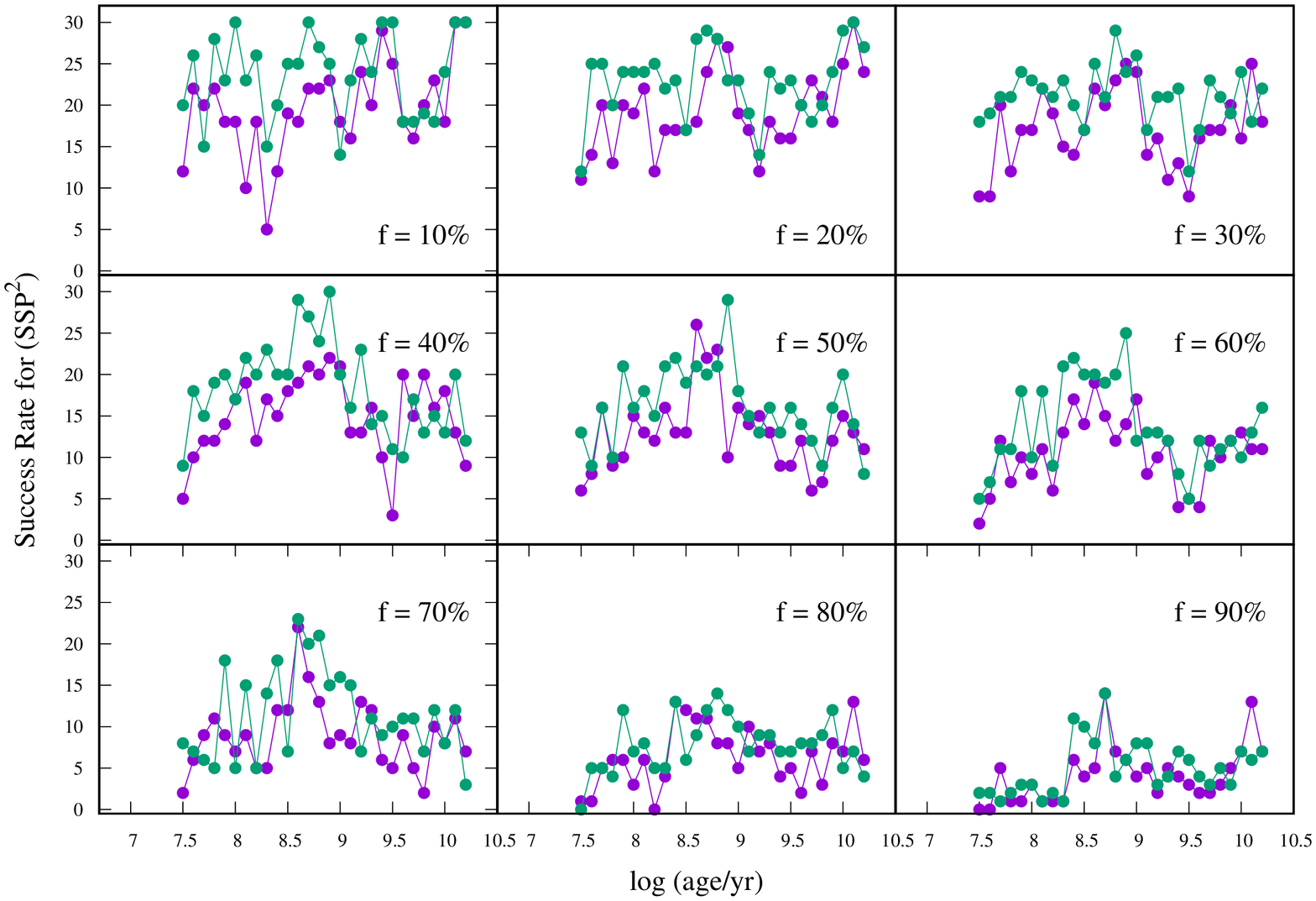}}
\caption{The number of successfully recovered ages (i.e., the number of times the derived age matches the real age $\pm$ 0.1 dex) as a function of real age for mock clusters with Age(SSP$_2$) -  Age(SSP$_1$) = 0.5. The top panel shows the results of Age(SSP$_1$) and the bottom panel shows the results of Age(SSP$_2$). The green dots represent results for S/N = 100 and the purple dots represent the results for S/N = 50. }
\label{Step5_Recovery}
\end{figure*}

\begin{figure*}
\resizebox{165mm}{!}{\includegraphics[angle=270]{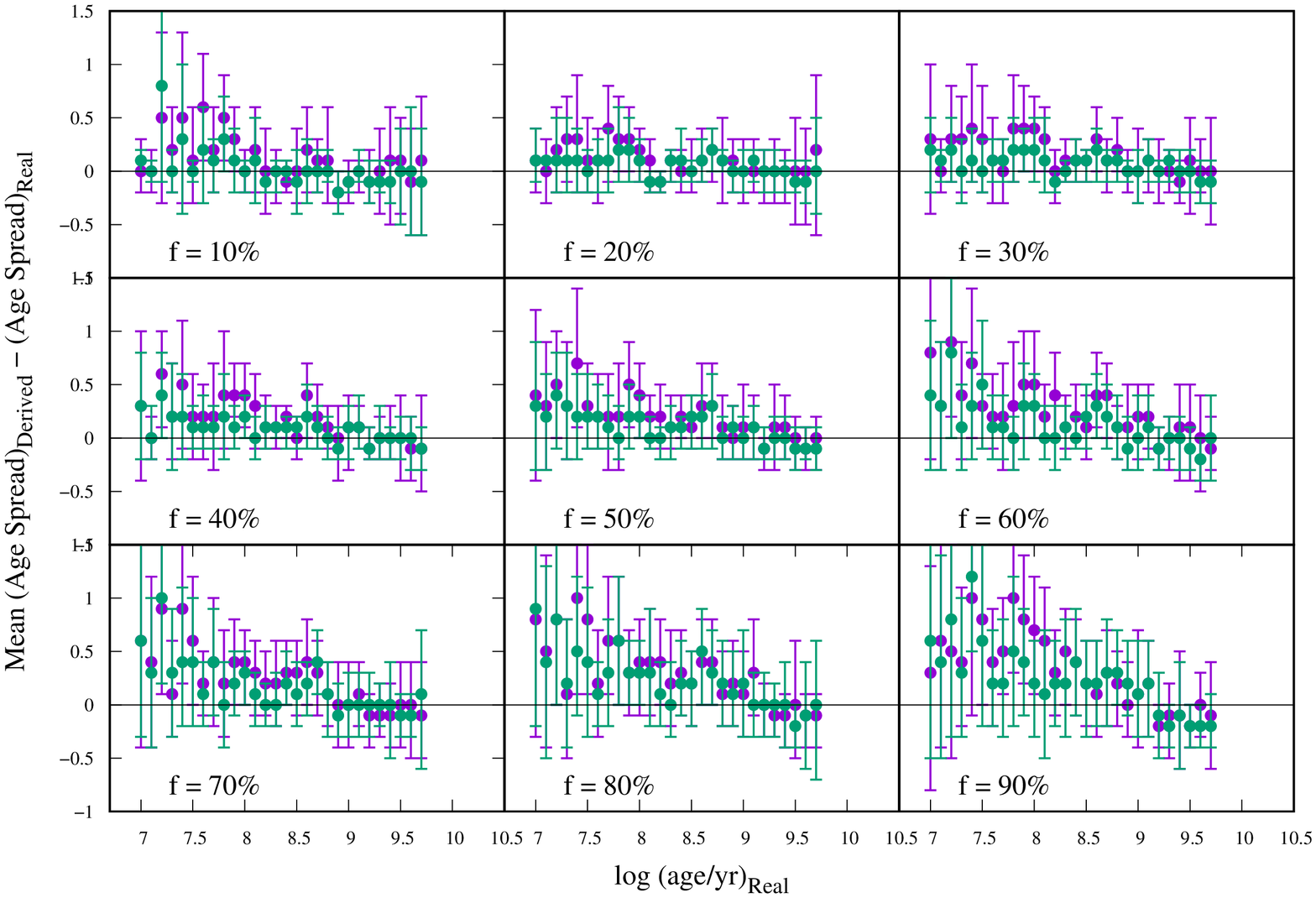}}
\resizebox{165mm}{!}{\includegraphics[angle=270]{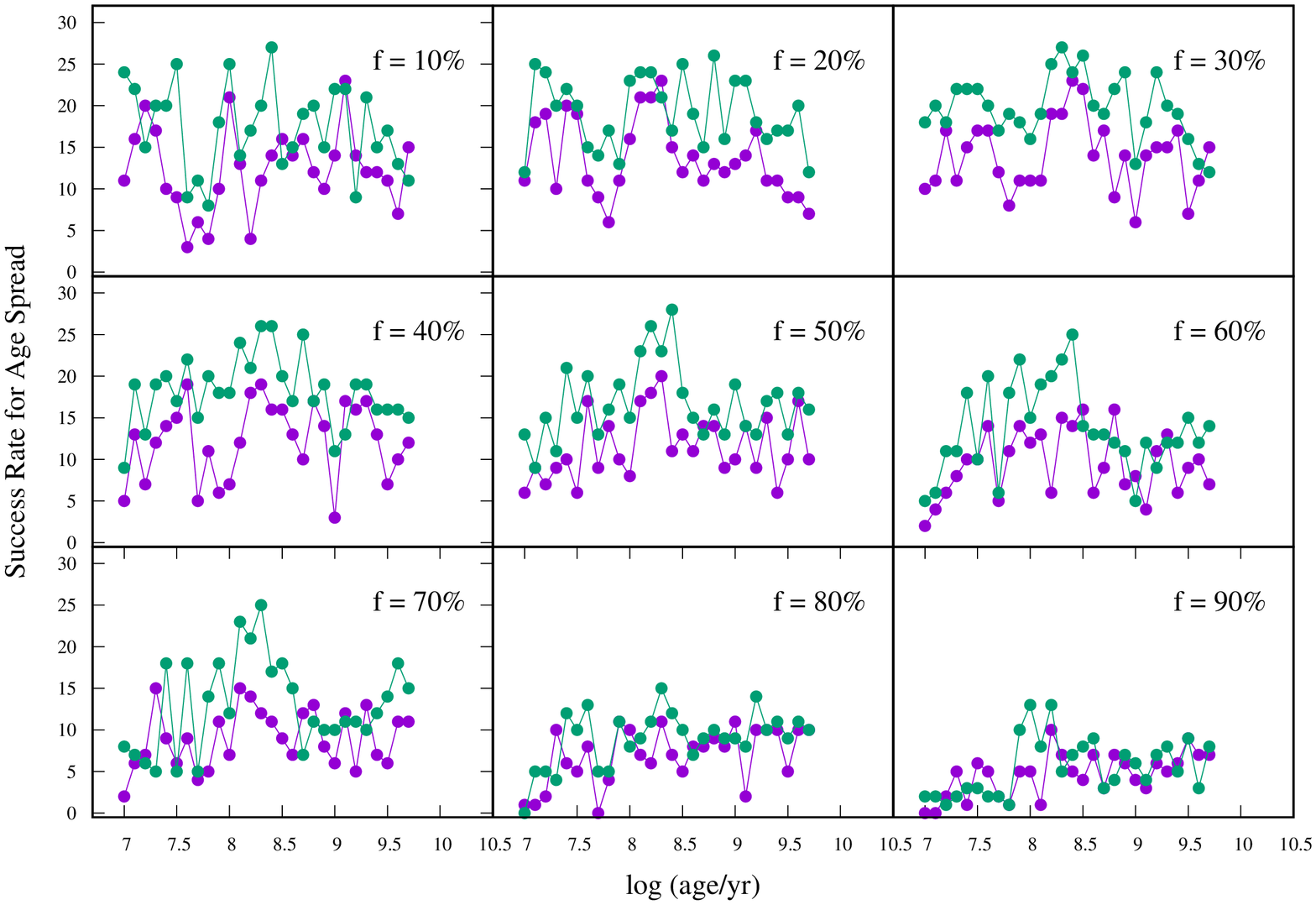}}
\caption{The same as Figure \ref{AgeGap_02}, but now for input $\Delta$\,(log\,(age/yr)) = 0.5}
\label{AgeGap_05}
\end{figure*}

In order to investigate the role of the size of the age gap on the derived parameters, we create a new set of 59,400 mock cluster spectra with larger age step:
\begin{equation} 
\log\, \mbox{(age(SSP$_{2}$))} = \log\, \mbox{(age(SSP$_{1}$))} + 0.5
\end{equation} 
 and repeat the analysis aiming to derive Age(SSP$_1$), Age(SSP$_2$) and $f$ from the full-spectrum fitting technique.
Figure \ref{Step5_Mean} shows that the mean of both Age(SSP$_1$) and Age(SSP$_2$) is closer to the real values for mock clusters with Age(SSP$_2$) -  Age(SSP$_1$) = 0.5 dex.
Despite the increase in the age gap, our results show that the full spectrum fitting technique still  overestimates Age(SSP$_2$) consistently for Age(SSP$_2$) < 8.5. 
Figure \ref{Step5_Recovery} shows the success rate as a function of input age.
For Age(SSP$_1$), the success rate depends on both the mass fraction $f$ and the input age. For mass fraction $f$ = 10, when SSP$_2$ is dominating, the success rate is very poor, however for mass fraction $f$ > 30 the success rate increases significantly, with drops seen around log (age/year) = 8.0 and log (age/year) > 9.4 due to the similarity of the SEDs in the age ranges where the success numbers are low.
When comparing the success rate of both Age(SSP$_1$) and Age(SSP$_2$) for mock clusters that have age gap = 0.5 dex with that of mock clusters that have age gap = 0.2 dex, we notice that the numbers are higher for age gap = 0.2. The reason is that by definition the success rate is the number of times the derived age matches the real age to within \textit{$\pm$ 0.1 dex}, hence smaller age gaps are closer to this precision limit. 
The mean of the derived age spread for an input age gap of 0.5 is shown in Figure \ref{AgeGap_05}. The results are better than for an age gap of 0.2, although the recovery rate of success to within 0.1 dex (lower panel) did not change significantly.

\section{Recovery of Mass Fraction}

\begin{figure*}
\resizebox{165mm}{!}{\includegraphics[angle=270]{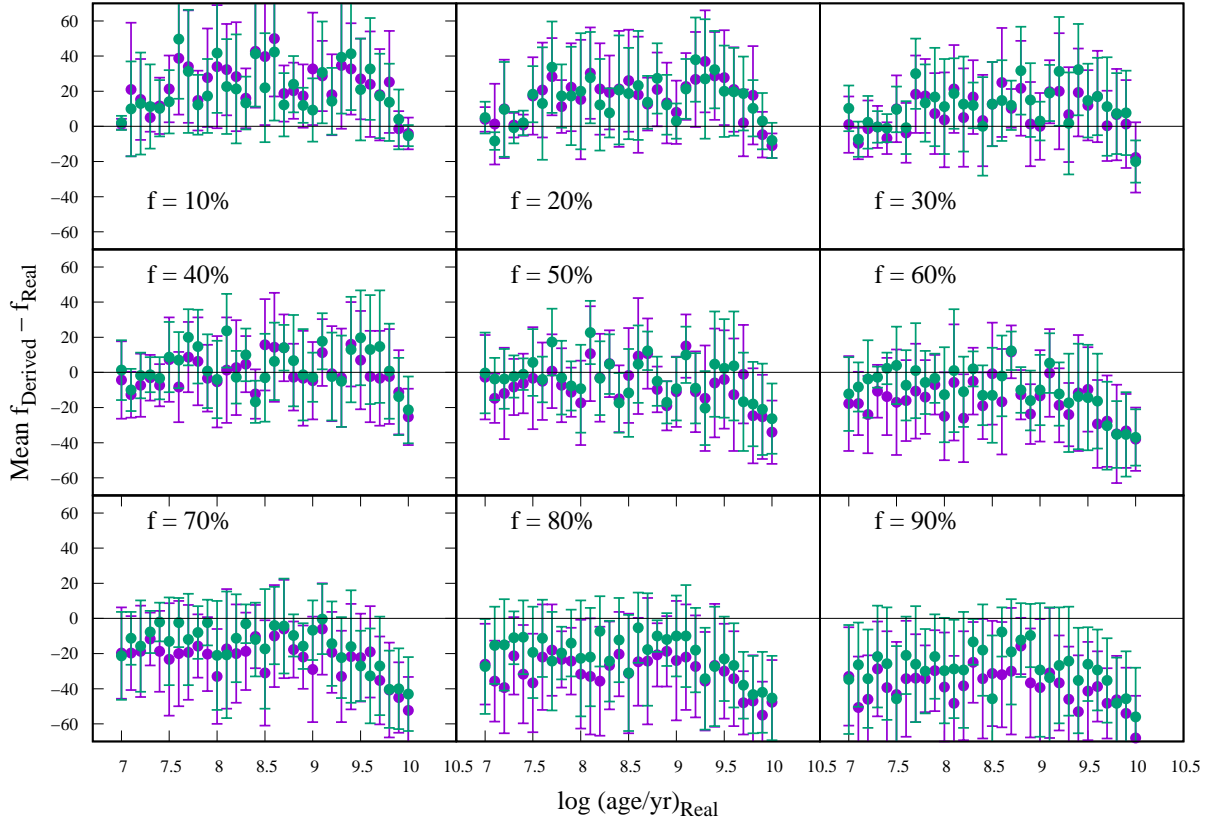}}
\resizebox{165mm}{!}{\includegraphics[angle=270]{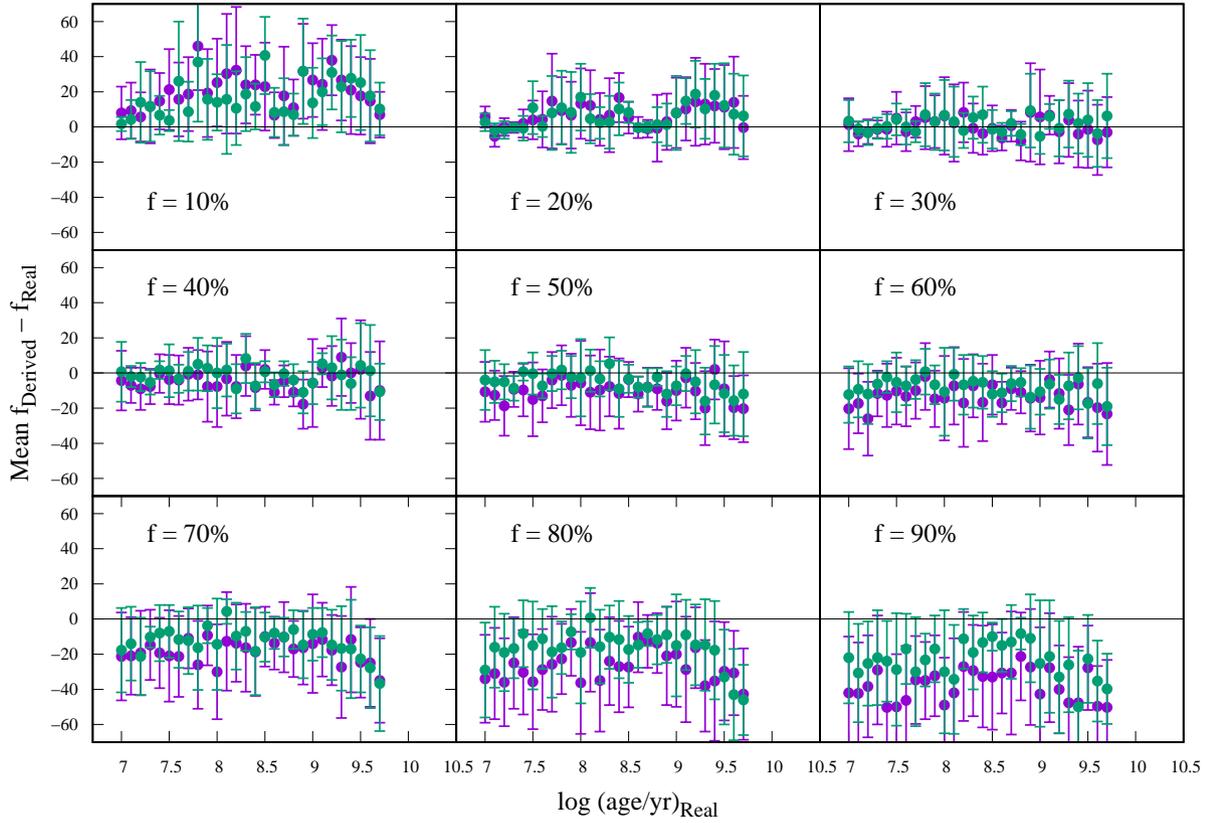}}
\caption{The mean of the derived mass fraction as a function of the input age. The top panel shows the results from $\Delta$ SSP = 0.2 and the lower panel shows the results from $\Delta$\,(log\,(age/yr)) = 0.5. The green dots represent results for S/N = 100 and the purple dots represent the results for S/N = 50. }
\label{Mass_Fraction}
\end{figure*}

Now we discuss the precision in recovering the mass fraction for the two cases analyzed in the previous sections (i.e., $\Delta$\,log\,(age/yr) = 0.2 and 0.5). Figure \ref{Mass_Fraction} shows the mean of the derived mass fraction as a function of the input age. Overall the derived mass fraction is overestimated for low mass fractions and underestimated for high mass fractions. The best results are achieved when the mass fraction is around 50\%.

\section{Summary}

We investigate the accuracy and precision of recovering possible age spreads within a star cluster using the full-spectrum fitting method. We analyze optical spectra of 118,800 mock star clusters with mass fractions from 10\% to 90\% for two age gaps (0.2 and 0.5 dex) in the age range 6.8 $<$ log\,(age/yr) $<$ 10.2. Random noise is added to the model spectra to achieve S/N ratios between 50 to 100 per wavelength pixel. We summarize our results in the following points:

A) The results obtained using the full-spectrum fitting method to derive the age of a star cluster that has an age spread of 0.2 dex, depend on the age and the mass fraction of each of the combined SSPs. If one SSP is dominating the spectrum, the full-spectrum fitting technique correctly predicts the age of the dominating SSP. When the contribution of the two SSPs is roughly equal, the full-spectrum fitting technique correctly predicts the average age value of the two SSPs [i.e., ((SSP$_1$)+(SSP$_2$))/2]. For all ages and mass fraction values, there is no significant dependence on the S/N value in the range 50\,--\,100.\\
B) When using the full-spectrum fitting method to derive combinations of ages with different mass fractions, the mean of the derived Age(SSP$_1$) matches the real Age(SSP$_1$) to within 0.1 dex up to ages around log\,(age/yr) = 9.5. The precision decreases for log (age/yr) $>$ 9.6 for all mass fractions or S/N values. This is because SSP SEDs are very similar in that age range, which makes it difficult to identify their correct age. \\
C) For young ages (log (age/yr) $\la$ 8.0) the full spectrum fitting technique tends to derive combinations of a young age with a significantly older age. This is because young populations (log (age/yr) $\la$ 7.0) produce significantly more flux than old populations, hence different combinations of young populations with old populations will have very similar $\chi^2$ values, causing the large uncertainty for the mean of the derived Age(SSP$_1$) and Age(SSP$_2$) as shown in Figure \ref{2SSPs_ss2}.\\
D) The success rate (defined as the number of times the derived parameter matches the input parameter $\pm$ 0.1 dex) peaks around log (age/year) = 7.2 and 9.0 and drops around 8.6 and $>$ 9.5 due to the similarity of SED shapes for those ages.\\
E) Increasing the age gap in the mock clusters improve the derived parameters. However, Age(SSP$_2$) is still overestimated for Age(SSP$_2$) < 8.5. \\
F) The derived mass fraction is overestimated for low mass fraction values and underestimated for high mass fraction values despite the age gap size. The best results are derived when the mass fraction is around 50\%.

\section*{Acknowledgments}
RA thanks the Space Telescope Science Institute for a sabbatical visitorship including travel and subsistence support as well as access to their science cluster computer facilities.
The initial analysis of this work was done using the Geospatial Analysis Center (GAC) computers at the American University of Sharjah. We acknowledge the support of GAC and IT staff. 
This work is based on work supported in part by the FRG17-R-06 and EFRG-18-SET-CAS-74 grants P.I., R.\ Asa'd from American University of Sharjah and Mohammed Bin Rashid Space Center (MBRSC) grant 201602.SS.AUS P.I. R.\ Asa'd at the American University of Sharjah.

\section*{Data availability}
The spectra of the mock star clusters and model SSPs created for this paper are available upon request from the corresponding author.

\bibliographystyle{mnras}
\bibliography{References}

\end{document}